\documentclass{aa}
\usepackage{graphicx}

\newcommand{\HI}{{\ion{H}{1}}}

\newcommand{\kms}{$\,$km$\,$s$^{-1}$}

\newcommand{\mJybeam}{mJy beam$^{-1}$}
\newcommand{\msun}{{$M_\odot$}}
\newcommand{\lsun}{{$L_\odot$}}

\def\HI{H{\,\small I}}

\def\emph#1{{\sl #1}}
\newcommand{\ltsima} {$\; \buildrel < \over \sim \;$}
\newcommand{\gtsima} {$\; \buildrel > \over \sim \;$}
\newcommand{\lta} {\lower.5ex\hbox{\ltsima}}
\newcommand{\gta} {\lower.5ex\hbox{\gtsima}}

\input{psfig}

\begin{document}

\title{B2~0648+27: a radio galaxy in a major merger.
\thanks{Based on observations  with the Westerbork Synthesis Radio Telescope
(WSRT) and the Very Large Array (VLA).}}

\titlerunning{\HI\ in the radio galaxy B2 0648+27}
\authorrunning{Morganti et al.}

\author{R. Morganti\inst{1},
 T.A. Oosterloo\inst{1}, A. Capetti\inst{2}, H.R. de Ruiter\inst{3,4},
 R. Fanti\inst{3,5}, P. Parma\inst{3} \\
 C.N. Tadhunter\inst{6} \and K.A. Wills\inst{6} }

\offprints{morganti@astron.nl}

\institute{Netherlands Foundation for Research in Astronomy, Postbus 2,
NL-7990 AA, Dwingeloo, The Netherlands
\and
Osservatorio Astronomico di Torino, Strada Osservatorio 25,
I-10025 Pino Torinese, Italy
\and
Istituto di Radioastronomia, Via Gobetti 101, I-40129, Bologna, Italy
\and
Osservatorio Astronomico di Bologna, Via Ranzani, 1, I-40127 Italy
\and
Dipartimento di Fisica dell'Universit{\`a} di Bologna, Via Irnerio 46,
I-40126 Bologna, Italy
\and
Dep. Physics and Astronomy,
University of Sheffield, Sheffield, S7 3RH, UK
}
\date{Received ...; accepted ...}

\abstract{We present WSRT  observations of the neutral hydrogen in
the nearby radio galaxy B2~0648+27. In emission, we detect a very
large amount of \HI\ ($M_{\rm HI} = 1.1 \cdot 10^{10}$ $M_\odot$) that
is distributed in a very extended disk, or ring-like structure, of
about 160 kpc in size. We also detect \HI\ absorption against the
central radio continuum component.  The detection of the
\HI, its structure and kinematics, give us key information for
building a possible evolutionary scenario. The characteristics of the
detected \HI\ are explained as the result of a {\sl major merger
event} that is likely to have occurred $\lta 10^9$ yr
ago. Interestinly, we find that, when observed in radio continuum at
higher resolution, this galaxy has a double lobed, steep spectrum
structure of about 1~kpc in size. Thus, despite its low radio power,
B2 0648+27 bears striking similarities with Compact Symmetric Objects,
i.e.\ objects believed to represent the early phase of radio galaxies
($\lta$ few thousand yrs old).  B2~0648+27 is one of the few nearby
radio galaxies where extended neutral hydrogen has been detected so
far. This, and other recent results, appear however to indicate that
nearby radio galaxies are more often gas rich than commonly assumed.
The phenomena described are likely to be much more common at high
redshift and galaxies like B2 0648+27 may provide valuable information
on the evolution of high redshift radio sources.  
}
\maketitle

\section{Mergers and the origin of radio galaxies}

\label{intro}

Galaxy mergers and interactions are often invoked as trigger of the
nuclear activity in radio galaxies. The large number of powerful radio
galaxies showing peculiar optical morphology (tails, bridges, shells
and dust-lanes; Smith \& Heckman 1989; Heckman et al.\ 1986), together
with the detection of large amounts of CO in a number of them
(Mazzarella et al.\ 1993, Evans et al.\ 1999a,b), can be taken as
evidence for this.  The increasing evidence that super-massive black
holes reside in the centres of many elliptical galaxies (e.g.\
Kormendy \& Richstone 1995) and the detection of potential fuel for
AGN (ionized gas and dust) in both radio loud and quiescent galaxies
(see e.g.\ Verdoes Kleijn et al.\ 2002a and references therein),
suggest that radio activity may occur (perhaps as a {\sl short phase})
in the life of {\sl many (or even all)} elliptical galaxies.

The nature of the triggering events remains, however, uncertain.  It
is not clear whether the activity is mainly triggered by {\sl major
mergers} between gas-rich galaxies or by {\sl minor accretion} events.
Perhaps either kind of merger plays an important role, as long as the
initial conditions are such that the gas will be able to reach the
very inner regions (on pc-like scales),  see e.g. Hibbard 1995, Wilson
1996.  Another uncertain aspect is at which stage of the merger the
onset of the radio activity takes place.

Evolutionary sequences linking merger-related objects have been
suggested by many authors. The case for a link between ultra-luminous
far-IR galaxies (ULIRG) and radio galaxies (e.g.\ Sanders \& Mirabel
1996, Evans et al.\ 1999a) is perhaps the strongest.  ULIRGs are
thought to be connected to gas rich mergers. A relation between
ULIRGs and radio galaxies would explain, e.g., the large amount of
molecular gas and dust detected in some radio galaxies and the
starburst component that is visible in the optical spectra of a number
of them  (Wills et al. 2002). In this
sequence, the radio activity is believed to appear late in the
evolution of the merger, when the molecular gas finally agglomerates
in the nuclear region of the galaxy (Evans et al.\ 1999a).

\begin{table}
\centering
\caption{Instrumental Parameters of the WSRT observations}
\label{tab1}
\begin{tabular}{lcc}
\hline\hline\\
Field Centre (J2000) & 06:52:02.5 27:27:39     \\
Date of the observations & 04Feb01  \\
Shortest spacing (m)     & 48 \\
Synthesized beam (arcsec)  & $64\times 27$ (p.a. 0.6$^\circ$)    \\
Number of channels       & 128 \\
Bandwidth (MHz)          & 10 \\  
Central frequency (MHz)          & 1363.8    \\
Velocity resolution (km/s) & 36  \\
(after Hanning)    & \\
rms noise in channel maps (mJy/beam) & 0.22 \\
\hline \\
\end{tabular}
\end{table}

\begin{figure*}  
\centerline{\psfig{figure=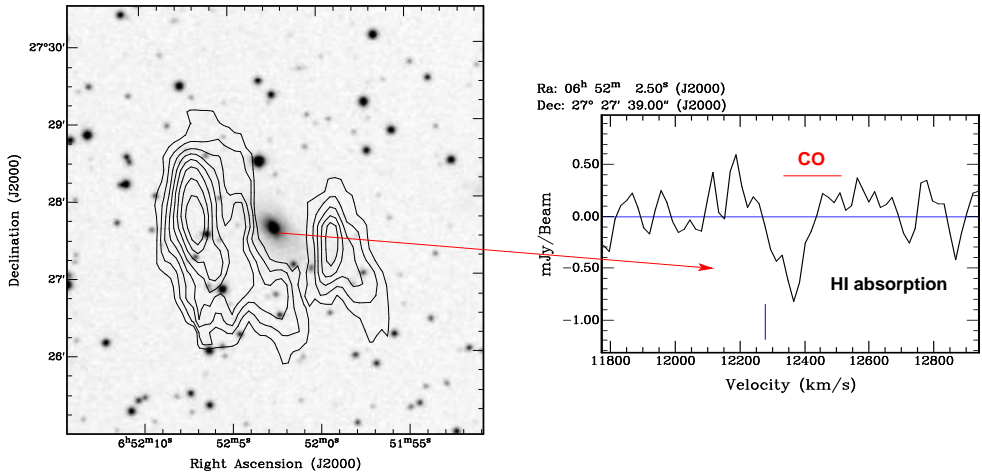,angle=0,width=13cm}}
\caption{{\sl Left} \HI\ total intensity contours  of 
the radio galaxy B2 0648+27  superimposed on to an optical image.
{\sl (Right)} \HI\ absorption profile, the optical 
systemic velocity is marked.
The range of the CO emission (from Mazzarella et al. 1993) is also indicated.
Contour levels: $4.3 \times 10^{19}$ to  $1.9 \times 10^{20}$ cm$^{-2}$ in steps of $1.8 \times 10^{19}$ cm$^{-2}$.}
\label{fig:prof}
\end{figure*}

The details of the overall evolution are, however, still quite vague.
In order to study these, a diagnostic of the conditions and kinematics
of the gas on large scales is important. This is because large scales
imply longer time scales for this gas to settle in a stable
configuration. The signatures and characteristics of the merger can,
therefore, be recognised over a longer period.  This is an advantage
over e.g.\ imaging studies of the centres of active and non-active
galaxies performed with HST (e.g. Verdoes-Kleijn 2002b and ref. therein,
Capetti et al. 2000). The dynamical time scales in these central
regions are very short and differences between active and non-active
galaxies can disappear quickly.

Observations of the neutral hydrogen can be a good tool to study the
evolution of mergers.  Through the amount of \HI\ (on large scales),
its morphology and kinematics, one can obtain a good idea about
what kind of galaxies are/were involved as well as about the age of
the merger. Large tails of neutral hydrogen signal that the merger is
relatively young. On the other hand, the large, regular gas disks found
in some elliptical galaxies indicate an older merger, where the gas has
had time to settle in such a disk (see Barnes 2002, Oosterloo et al.\ 2002a,
Morganti et al.\ 1997, van Gorkom \& Schiminovich 1997).

In order to obtain more information on the AGN-phase in mergers and to
investigate the connection with ULIRGs, gas-rich and normal elliptical
galaxies, a thorough study of the \HI\ properties of radio galaxies is
essential for radio-loud galaxies. A good candidate for such a study
is B2~0648+27, a radio galaxy that has a number of characteristics
that indicates that it may be related to ULIRGs or ULIRG-like
galaxies.  Here we present the results of
\HI\ observations carried out with the Westerbork Synthesis Radio
Telescope (WSRT).

\section{B2~0648+27 and the WSRT observations}

B2~0648+27 is a compact radio source ($\log {\rm P} = 24.02$ W/Hz at
1.4 GHz\footnote{For $H_\circ = 50$ \kms Mpc$^{-1}$ and $q = 0.0$, 1
arcsec = 1.1 kpc}) hosted by a luminous elliptical galaxy ($M_B =
-22.94$ and $\log L_B = 11.34$ $ L_\odot$).  This galaxy has a
particularly interesting interstellar medium (ISM).  It contains $\sim
6 \times 10^9$ $M_\odot$ of molecular hydrogen and it is also an IRAS
source ($L_{\rm FIR}\sim 2\times 10^{11}$ $L_\odot$, Mazzarella et
al.\ 1993). It shows warm far-infrared colours and is classified as a
60-$\mu$ peaker (Vader et al.\ 1993, Heisler et al.\ 1998), likely
resulting from the large amount of dust heated by the nuclear activity
but also by the star formation (see below).  The estimated mass of the
dust is $\log M_{\rm dust}/M_\odot = 6.48$. The central regions have
been observed with HST and the dust it is found to be distributed both
in a regular disk and in patches (Capetti et al.\ 2000, de Ruiter et
al.\ 2002).  The presence of a young stellar population has been
determined from optical spectra (Ebneter 1989). The clear deviation
from the far-IR/radio correlation (Heisler et al.\ 1998) and the warm
IR colours (i.e.\ the ratio 25/60 $\mu$) typical of AGN, confirm the
classification as radio galaxy.

The redshift of B2~0648+27 derived from optical lines is $12270\pm 120$
\kms\ ($z = 0.0409 \pm 0.0004$, Falco et al.\ 2000) while 12420 \kms\ ($z =
0.041$) was derived from the CO detection (Mazzarella et al.\ 1993).

B2~0648+27 was observed using the WSRT at the frequency of the
redshifted \HI\ line for 12 hours. The parameters of the observations
are summarized in Table~1.  The data were calibrated using the MIRIAD
package (Sault et al.\ 1995).  The final cube was made with
robust--Briggs' weighting equal to 1 (Briggs 1995).  The restoring
beam is $64 \times 27$ arcsec (p.a.\ 0.6 degrees) and the rms noise is
0.22 \mJybeam.  The velocity resolution is, after the Hanning
smoothing we applied, 36 \kms.  At the resolution of our observations,
the radio continuum of this galaxy appears unresolved (but see Sec. 4
for the high resolution continuum image).

\section{The neutral hydrogen in B2~0648+27}

In B2~0648+27 we detect \HI\ both in {\sl emission} and in {\sl
absorption}. The emission comes from a large amount of \HI\ in a
structure around the galaxy, while the absorption is detected against
the unresolved central radio continuum component.

\subsection{\HI\ emission}\label{emission}

The striking property of the \HI\ emission is the large amount that is
detected.  The total \HI\ mass is estimated to be $M_{\rm HI} = 1.1
\times 10^{10}$ \msun.  This is an unusually large amount for an
elliptical galaxy.  The optical luminosity of the host galaxy is also
high (the RC3 lists $B_T^0=14.03$ which implies $L_B = 2.2 \times
10^{11}$ \lsun\ and therefore the relative \HI\ content is only $M_{\rm
HI}/L_B$ = 0.05 $M_{\sun}/L_{B,\sun}$.

Figure 1 shows the \HI\ total intensity superimposed on an optical
image taken from the Digital Sky Survey.  The \HI\ emission is very
extended, the diameter is about 2.5 arcmin corresponding to $\sim 160$
kpc. The surface density of the \HI\ is quite low, the peak of the
emission is only 0.8 $M_\odot$ pc$^{-2}$. Therefore, in order to get
detailed information on the morphology and kinematics of the \HI\
deeper observations will be required.  However, from the available
data we can infer some important characteristics of the \HI. Figure 1
shows that the brightness distribution is asymmetric and is much
brighter on the E side. The neutral gas shows a smooth velocity
gradient that brackets the systemic velocity and this could indicate
that rotation is the main component of the observed
kinematics. However, the range of velocities is asymmetric with
respect to those of the CO which, if we assume that the CO is located
in the central regions, indicates that most of the \HI\ is not in an
equilibrium configuration.  The position-velocity plots (Figs. 2 and 3)
suggest that the \HI\ is perhaps not in a disk given that the
\HI\ appears almost as a linear feature in these plots. Such a
shape is more characteristic of of an incomplete ring or annulus (or even a
tail-like structure). The apparent lack of
\HI\ emission at the position of the optical galaxy (Fig. 1 and 2) could be 
partly due to the presence of \HI\ absorption against the
compact radio source (see below).
Higher quality  data are needed to be able to say more about 
the structure of the \HI.

%The fact that the range in velocity of the \HI\ (about 400 \kms) is also
%much broader than what is observed in CO (which is about 220 \kms)
%suggest that the \HI\ and the CO are not co-spatial. 

\subsection{HI absorption}

We also detected \HI\ in absorption against the unresolved central continuum
source  of the radio galaxy. The \HI\ absorption profile is shown in
Fig.~1.  The peak absorption is $\sim 0.84$ mJy beam$^{-1}$. From the
continuum image derived using the data from the line-free channels, we
derive a continuum flux  of $\sim 147$ mJy, comparable to the 156 mJy
obtained from VLA data at 1.4 GHz by Fanti et al.\ 1987.
This gives an optical depth of 0.6\%.

The FWHM of the absorption profile is $\sim 100$ \kms\ and the central
velocity ($\sim 12370$ \kms). The FWHM of the CO appears to be broader
(FWHM$=220$ \kms, Mazzarella et al.\ 1993) than the \HI.  The column
density is $\sim 1.1 \times 10^{20}$ cm$^{-2}$ for $T_{\rm spin}=100$
K. This is similar to what found in some other radio galaxies (see e.g.\
Morganti et al.\ 2001, Pihlstr\"om 2002) when sensitive observations
are available. It is also very similar to the column densities of the
\HI\ seen in emission.  The \HI\ seen in absorption is likely to be part of
the same system as the gas detected in emission but  happens
to be in front of the radio source, hence it is gas at large
radius. Thus, the difference in width between the \HI\ and the CO likely
reflects the difference in location of these two gas components, with
the CO closer to the nucleus compared to the neutral hydrogen.
Unfortunately, no image of the distribution of the CO emission is
available.

\begin{figure}  
\centerline{\psfig{figure=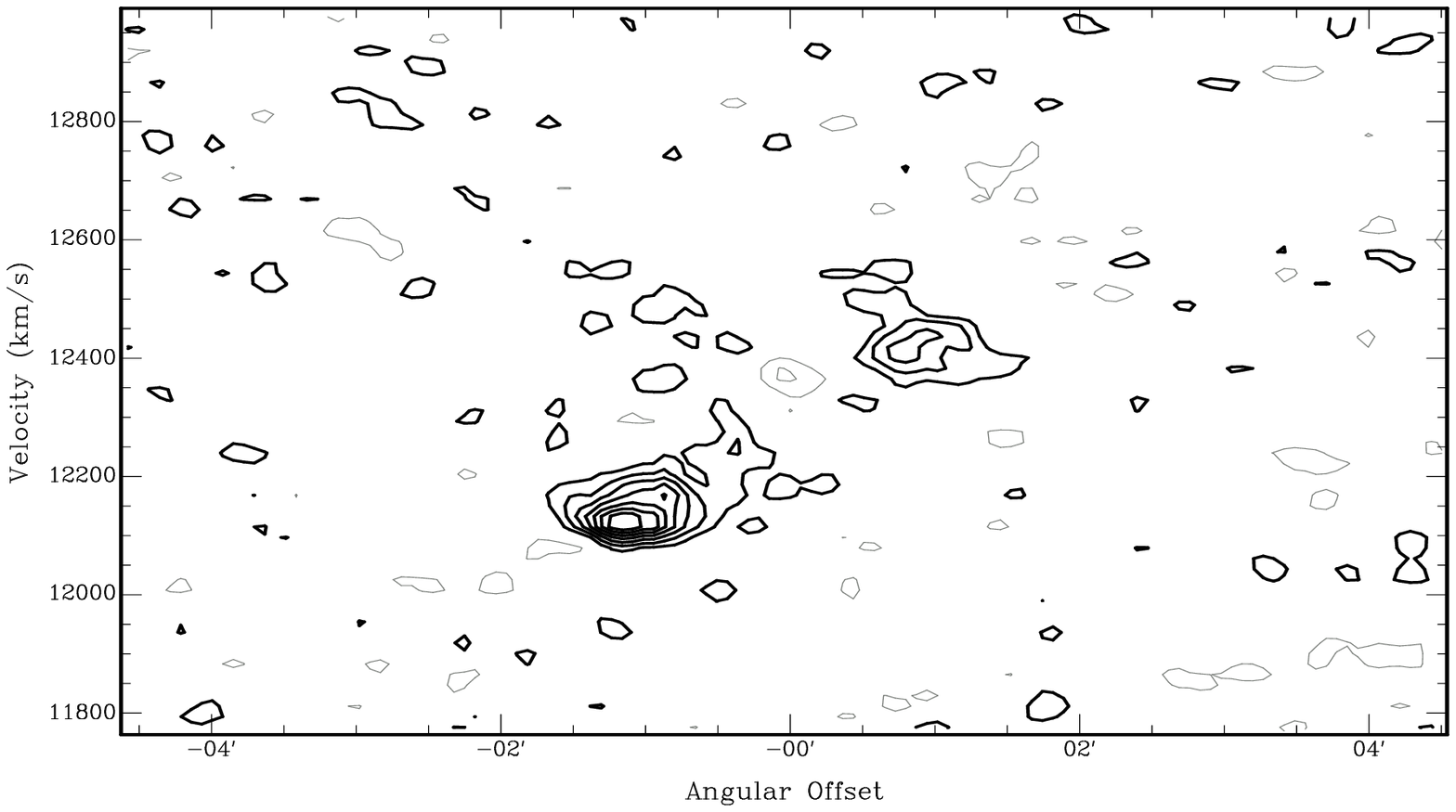,angle=0,width=7cm}}
\caption{Position-velocity  plot along the major axis (p.a.$ = 250^\circ$). The contour levels are -0.8, -0.4, 0.4 to 2.8 mJy beam$^{-1}$ in step of 0.4 mJy beam$^{-1}$.}

%\begin{figure}  

\centerline{\psfig{figure=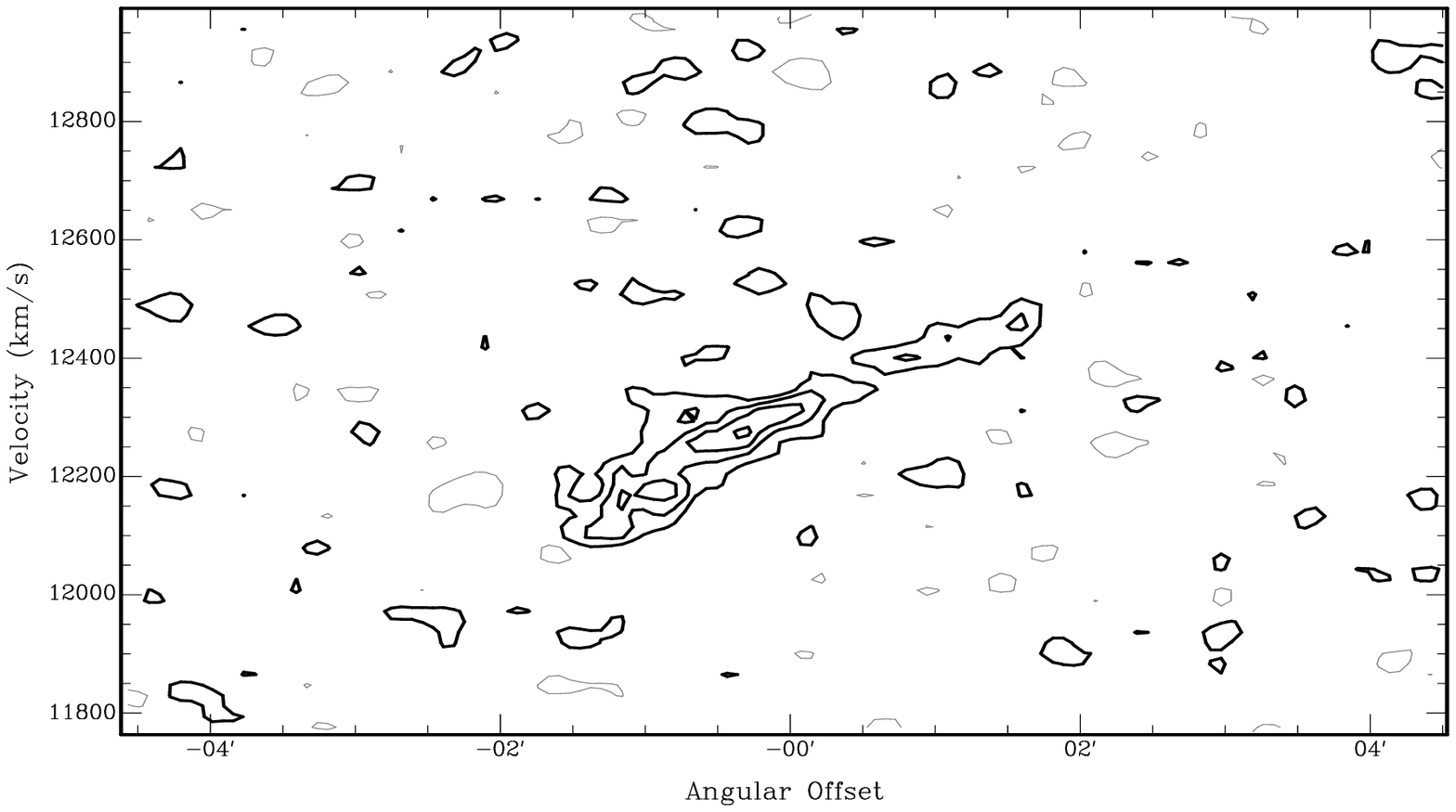,angle=0,width=7cm}}
\caption{Position-velocity  plot along the p.a. of the major axis (p.a.=250$^\circ$) but $\sim 40$ arcsec offset  to the south. The contour levels are -0.8, -0.4, 0.4 to 1.6 mJy beam$^{-1}$ in step of 0.4 mJy beam$^{-1}$.}
\end{figure}

\subsection{Other objects in the field }\label{other}

Three other objects are detected in \HI\ emission in the field of
B2~0648+27.  Their total \HI\ images are presented in Fig.~4.  The three
galaxies are separated from each other between 300 and 800~kpc and they
form a small group, situated about 27 arcmin (or 1.8 Mpc) east of
B2~0648+27.  Moreover, the average redshift of the galaxies is about
1000 \kms different from B2 0648+27.  Ongoing or recent interactions
between these galaxies and B2~0648+27 can therefore be excluded, because
the group is quite distant from B2 0648+27 while it also appears to be
at a somewhat lower redshift. 

A summary of their characteristics is given in Table~1.  The brightest
one (J0652+2721) is identified with a 2MASS galaxy (2MASXi
J0652557+272157) and the systemic velocity for this galaxy is $V_{\rm
hel} \sim 11320$
\kms.   The \HI\ profile
has a FWHM of only $\sim 50$ \kms, probably also due to the fact that
the emission is situated at the very edge of our observing band and we
could be missing some emission coming from this object.  We also
detect
\HI\ in coincidence of an anonymous low-surface brightness galaxy, J0652+2719,
with a systemic velocity of $V_{\rm hel} \sim 11620$ \kms\ and the \HI\
profile has a FWHM $\sim 80$ \kms.

The third \HI\ detection is coincident with an anonymous galaxy,
J0652+271509, with a velocity of $V_{\rm hel} \sim 11455$ and a FWHM of
$\sim 60$ \kms.

Because of their large distance from the field center, the \HI\  flux of these
galaxies is very much affected by the attenuation of the  primary
beam. Therefore any estimate of the \HI\ mass would be very uncertain.

\begin{table}
\centering
\caption{.}
\label{tab1}
\begin{tabular}{lccc}
\hline\hline\\
{\bf Name } &    $\alpha$ & $\delta$  & $V_{\rm hel}$   \\
            &     (J2000) & (J2000)   &   (\kms)        \\
\hline    
J0652+2721 &   $06^{\rm h}52^{\rm m}56^{\rm s}$  &  
$27^\circ 21^\prime 57^{\prime\prime}$ & 11320    \\
J0652+2719 &   $06^{\rm h}52^{\rm m}45^{\rm s}$  &  $27^\circ 19^\prime 28^{\prime\prime}$ & 11620 \\
J0652+2709 & $06^{\rm h}52^{\rm m}48^{\rm s}$ & $27^\circ 09^\prime
45^{\prime\prime}$ & 11455\\ 
\hline \\
\end{tabular}
\end{table}

\begin{figure}  
\vskip 5cm
%\centerline{\psfig{figure=Morganti.fig4.ps,angle=0,width=8cm}}
\caption{Total \HI\ intensity (contours) of the three galaxies  detected in
  the field of B2~0648+27 superimposed on an optical images taken from
  te Digital Sky Survey. Contour levels: $4.3 \times 10^{19}$ to $1.3
  \times 10^{20}$ cm$^{-2}$ in steps of $1.8 \times 10^{19}$
  cm$^{-2}$.}
\end{figure}
 
\section{VLA 8~GHz observations}

VLA observations at 8~GHz using the A-array configuration where obtained
on 03-May-2002 in order to understand the sub-arcsec structure of this
galaxy. The final image is shown in Fig.~5. The restoring beam is
$0.2 \times 0.19$ arcsec (p.a.\ = --12$^\circ$) and the rms noise 26 $\mu$Jy
beam$^{-1}$. The peak of the continuum emission is 10.7 mJy.

At this sub-arcsec resolution, B2~0648+27 is resolved into a
double-lobed structure.  This is consistent with the fact that the
overal radio spectrum of the source is steep ($\alpha^5_{0.4}=0.6$,
$S\sim \nu^{-\alpha}$), similar to that of large scale radio sources,
and not typical of radio cores.  The overall structure is about 1 kpc
in size. The radio morphology of B2~0648+27 is therefore similar to
the so-called Compact Steep Spectrum sources and Compact Symmetric
Objects (see O'Dea 1998 for a review), objects that are believed to
have a small size, not due to beaming effects, but because of the
young age of their radio emission ($\lta$ few thousand yrs old).
B2~0648+27 could be, therefore, a low-power example of this group of
sources.  The possibility that the radio emission in B2~0648+27
started not so long ago is interesting in the light of the time scales
of the merger that can be derived from the HI and the evolutionary
sequence that will be discussed in Sec. 5.3.

\begin{figure} 
\centerline{\psfig{figure=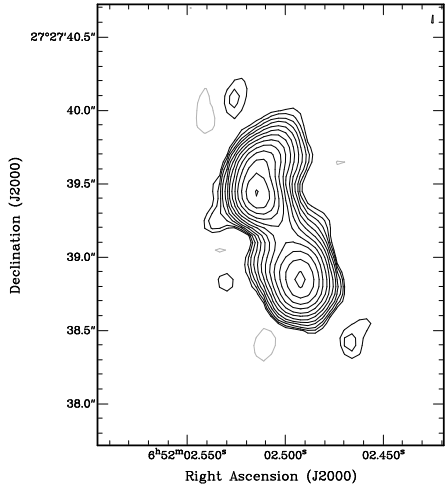,angle=0,width=7.5cm}}
\caption{VLA image of B2 0648+27 at 8~GHz. The contour levels are 
-0.08, 0.08 to 11 mJy beam$^{-1}$  increasing with a  factor 1.5. }
\end{figure}

\section{A radio galaxy in a major merger}\label{discuss}

\subsection{Origin of the \HI}

The main result of this study is that in the radio galaxy B2~0648+27,
\HI\ has been detected not only in absorption, as often is the case in
radio galaxies, but also in emission.  The amount of gas detected is
large ($\sim 10^{10}$ \msun) and distributed in a very extended
structure ($\sim 160$ kpc in diameter) of which the kinematics suggest
that it is not entirely in equilibrium.  Large amounts of \HI,
distributed over a large region, have now been found in a number of
nearby radio-quiet elliptical galaxies, sometimes, but not always,
distributed in a relatively regular disk-like structure (see e.g.\
Morganti et al.\ 1997, van Gorkom \& Schiminovich 1997, Oosterloo et
al.\ 2002b and references therein).   In these systems the neutral
hydrogen is explained as originating from a major-merger event,
involving at least one gas-rich, disk galaxy, that occurred, in some
cases, more than $10^9$ yr ago.  This appears also to explain the
characteristics of the detected \HI\ in B2~0648+27.

Extended \HI\ in emission was also found by Lim \& Ho (1999)
associated with three quasars host galaxies.  The
\HI\ detected there exhibit ongoing or remnant tidal \HI\ disruptions
possibly tracing galactic encounters or mergers.  These objects are
perhaps in a early stage of their evolution.  Unlike B2~0648+27 they
are radio-quiet objects. Also in these cases the amount of
\HI\ mass is quite high (between 0.5 and $2.5 \times 10^{10}$
M$_{\odot}$). 

The structure of the \HI\ and the assymetric density
distribution suggest that in B2 0648+27 the \HI\ is unlikely to be in
a settled configuration.  A rough indication of the age of the merger
is $R/V$ where $R$ is the size of the system and $V$ a characteristic
velocity. For B2 0648+27 we find $R/V \simeq 3\cdot 10^8$ yr.

It is interesting to note that in B2~0648+27 the column density of the
\HI\ detected in emission is relatively low as in the other gas-rich
elliptical galaxies mentioned above (see e.g.\ Oosterloo et al.\
2002a).  The peak of the \HI\ surface density is only $\sim 0.8$
$M_{\sun}$pc$^{-2}$, well below that found in the inner regions of
most spiral disks.  Because of this low surface density, no
significant star formation is, at present, occurring in the regions
coincident with the \HI. The galaxy can, therefore, remain gas rich
for a very long period.

\begin{figure*}
\vskip 5cm
%\centerline{\psfig{figure=Morganti.fig6.ps,angle=0,width=18cm}}
\caption{Possible evolutionary sequence linking gas-rich mergers with radio
galaxies and gas-rich ellipticals (described in text). The images of
the Antennas and NGC~7252 have been taken from Hibbard \& van Gorkom
(2001), B2~0648+27 from this paper, NGC~5266 from the data presented
in Morganti et al. (1997).}
\end{figure*}

\subsection{Gas disks and radio activity}

The interesting element, compared to many other gas rich elliptical
galaxies, is that B2~0648+27 is also a relatively strong radio source. 
With the sensitivity of present day radio telescopes it is very
difficult to detect neutral hydrogen in emission at the typical distance
of radio galaxies.  \HI\ in emission is detected only in an handful of
very nearby radio galaxies, for example IC~1459 (Oosterloo et al.\ 1999,
NGC~4278 (Raimond et al.\ 1981, Lees 1994), NGC~1052 (van Gorkom et al.\
1986), PKS B1718-649 (V\'eron-Cetty et al.  1995) and Centaurus~A (van
Gorkom et al.\ 1990, Schiminovich et al.\ 1994).  With the exception of
the last two objects, these radio sources are at the lower end of the
distribution of radio power for radio galaxies and they are more than an
order of magnitude less powerful than B2~0648+27.   Among the
objects listed above, the one that shows characteristics very similar to
B2~0648+27 is the southern radio galaxy PKS~B1718-649.  This is a
compact steep spectrum source (Tingay et al.  1997) around which a large
($\sim 180$ kpc diameter) \HI\ disk of $\sim 3.1 \times 10^{10}
M_{\odot}$ has been found (V\'eron-Cetty et al.  1995). 

In addition to this, by using \HI\ absorption against the radio lobes
(i.e.  tens of kpc in size), extended neutral hydrogen has been detected
in the radio galaxy Coma~A (Morganti et al.\ 2002a).  In this object at
least $10^9$ $M_{\sun}$ of \HI\ have been detected in a disk-like
structure of at least 60 kpc in diameter.  Other examples have also been
recently found where the \HI\ absorption is not limited to the nuclear
region but extend tens of kpc, e.g, 3C~433 (Morganti et al.\ 2002b) and
3C~234 (Pihlstr\"om 2001) although they have not yet been studied in
detail.  Thus, the case of B2~0648+27, together with these other recent
results, indicates that extended structures of neutral hydrogen in radio
galaxies may actually be more common than thought so far.  A systematic
study of these systems is now essential. 

Although these objects may represent only a relatively small
subset in the group of  radio galaxies, they may have an extra relevance
as link to high-z objects.  Extended \HI\ absorption (observed against
the Ly$\alpha$ emission) has been found in a high fraction of high-$z$
radio galaxies (van Ojik et al.\ 1997). This is considered an
indication that high-$z$ radio galaxies are located in dense
environments and is a diagnostic for the effects of radio jet
propagation in this dense medium.  Moreover, a low surface brightness
Ly$\alpha$ halo with quiescent kinematics has also been found in the
case of the distant radio galaxy USS~0828+193 (Villar-Mart\'in et
al. 2002). 

Although these phenomena may be occurring more frequently and more
efficiently at high redshifts, in the \HI-rich, low redshift radio
galaxies we may witnessing a similar situation.  Indeed,
Villar-Mart\'in et al. 2002 indicate the possibility 
that the low surface brightness  Ly$\alpha$ halo in USS~0828+193 is
the progenitor of the \HI\ discs found in low redshift galaxies.

Finally, as mentioned above (Sec. 4), B2~0648+27 could be a young radio
source. For this and other characteristics (high far-IR luminosity,
\HI\ absorption, young stellar population component), B2~0648+27 bears
a strong similarity with two other radio galaxies, namely PKS~1547--79
and 4C~12.50 (see also Sec. 5.3). The properties of these sources have
been explained by that they are young sources still enshrouded in a
cocoon of material left over from the event which triggered the
nuclear activity. The radio source, during its evolution, is sweeping
away large amounts of obscuring material present along the radio axis.
In PKS~1547--79 and 4C~12.50 the interaction between the radio plasma
and the ISM is evidenced by the presence of two redshift systems (see
Tadhunter et al.\ 2001 and Morganti et al. 2003) for a detailed
discussion).  Good quality optical spectra are necessary to
investigate whether B2~0648+27 also shares these characteristics.
Unfortunately, given their higher redshift, the search for extended
neutral hydrogen is particularly difficult for PKS~1547--79 and
4C~12.50.  The finding of \HI\ in B2~0648+27 confirms, nevertheless,
the trend of an higher probability of detecting neutral hydrogen in
starburst/far-IR bright radio galaxies as found by Morganti et al.\
(2001). Although B2~0648+27 is actually the first case where \HI\ is
detected also in emission, this result can be taken as a further
indication that the interstellar medium is indeed, on average, richer
in these objects than in other radio galaxies.

\subsection{The merger and the order-of-events}

The relevance of major vs minor mergers as trigger for the radio
activity and the time of the onset of this activity in the life of a
galaxy, are not yet well understood. Major mergers have been proposed
to explain, e.g., the large amount of CO observed in some radio
galaxies.  The spread in the CO properties
observed in radio galaxies (Mazzarella et al.\ 1993, Evans et al.\
1999a,b) is used to argue that {\sl the radio activity appears late in
the life of the merger and a long time after the start of the
starburst phase} (as late as it takes for the molecular gas to
agglomerate in the very nuclear regions).  Depending on such a delay,
the consumption of the left-over gas (both molecular and neutral) by
extended star formation could be well underway even before the radio
activity starts and this will explain the large range in CO values
(see the case of 4C~12.50, Evans et al.\ 1999a).  Lim et al.\ (2000),
however, have argued that the fact that most of the radio galaxies lie
very close to the fundamental plane (unlike merging systems), is
better explained  by cannibalizing of smaller gas-rich galaxies by
the pre-existing galaxy that therefore retains its gross morphological
and dynamical properties.

The study of the stellar population is also giving a complex view of
what may be the trigger mechanism of radio activity and that we are
probably dealing with not a single type of merger as trigger of the
activity (Wills et al.\ 2002). A young stellar population component has
now been detected in about 30\% of the observed radio galaxies (Wills
et al.\ 2002, Tadhunter et al.\ 2002) and in some of them the mass of
this component is  large, suggesting its origin to be related to a
major merger event.  If this is the case,  the merger is
estimated to be relatively old ($>0.1$ Gyr, see e.g.\ Tadhunter et al.\ 2002,
Wills et al.\ 2002), supporting the idea that al least in those objects
the radio activity starts in these galaxies quite late after the
merger.

The case of B2~0648+27 has shown as the neutral hydrogen can be a good tool to 
answer some of the above open questions
and derive key information, not only about the origin,
but also about the evolution of the system and the time-scales
involved.  In the case of B2~0648+27, the likely origin of the system
is a major merger. This supports the idea of a link between, at least some, radio
galaxies and ULIRGs, as already suggested by the characteristics of the
ISM in this galaxy. The prototype of this class of objects is the radio
galaxy 4C~12.50 (Evans et al. 1999a) and the similarities between the
two objects have been already mentioned above (Sec. 5.2).

Although this could be the case for only a limited group of radio
galaxies, nevertheless we can attempt, using the information obtained
by the \HI, to include B2~0648+27 in the evolutionary sequence already
proposed for other gas-rich systems (see e.g.\ Hibbard
\& van Gorkom 1996, Wilson 1996, Mihos \& Hernquist 1994, Morganti et
al.\ 1997) This is presented in Fig.~6.

The Antennae (left) illustrates an ongoing merger where two galaxies
are still identifiable and where a starburst is occurring (Hibbard et
al.\ 2001).  Part of the \HI\ is found in two long tidal tails.  NGC
7252 (2\( ^{\rm nd}
\) from left) represents a somewhat later stage where the central
remnant has already more or less taken the shape of an elliptical
galaxy (Hibbard et al. 1996).  The \HI\ is mainly in large tails at
large radii, while the gas in the centre, where much star formation is
still occurring, is mainly molecular.  NGC 7252 is a strong emitter in
the far infrared.  After an intermediate stage with AGN activity (B2
0648+27, see below), the final stage (NGC 5266; 4\( ^{\rm th} \)
panel), shows a galaxy that has become a genuine early-type galaxy
(Morganti et al.\ 1997).  The \HI\ is falling back from the tails to
the galaxy and is in the process of forming a large disk or ring-like
structure of low surface density.  In NGC 5266, in fact, two such
systems, that are orthogonal to each other, are observed (Morganti et
al.\ 1997). In the outer regions \HI\ filaments are still visible.
Star formation is occurring at a much reduced rate and no AGN is
detected.

The new \HI\ data allow to place B2~0648+27 on the evolutionary
sequence illustrated in Fig.~6.  The, admittedly very rough,
estimate of the age of the merger, based on the size and the
kinematics of the \HI, is several times $10^8$ yr. This suggests that
this galaxy can be placed between NGC~7252 and NGC~5266. Considering
that the timescale over which AGN activity occurs is at least an order
of magnitude smaller, in this object the AGN activity occurred at a
late stage of the merging process.  This is particularly interesting
when considering that the radio emission, indeed, appears to be relatively
recent in B2~0648+27. This is therefore consistent with the fact that
the \HI\ has not yet completely settled and therefore that the merger
is not too old.

A final word of caution is that the first-order kind of scenario
proposed above will have to be further investigated using different
kind of observations. In particular, it will be important to derive the age of
the young stellar population from optical spectroscopy, while deep optical
images will have to show whether indeed a stellar conterpart of the
\HI\ structure is present (as found in NGC~5266 and NGC 2865
(Morganti et al. 1997 and Schiminovich et al. 1995 respectively).

\section{Summary}

We have presented \HI\ observations of the radio galaxy B2~0648+27
where \HI\ both in absorption and in emission has been detected.  The
neutral hydrogen is distributed in a large, disk or ring-like
structure of $M_{\rm HI} = 1.1 \cdot 10^{10}$ $M_\odot$ in mass and
about 160 kpc in diameter.  This result, together with what recently
found in other objects, indicates that extended structures of neutral
hydrogen in radio galaxies may actually be more common than thought so
far.  The \HI\ can be used to study the likely origin of these systems
and the time scales involved. In the case of B2~0648+27, the large
amount of neutral gas suggests a major-merger event.  From the
kinematics and the size of the \HI\ we estimated the age of the merger
to be several times $10^8$ yrs.  Gas-rich, low redshift systems like
B2~0648+27 could be the nearby examples of phenomena that are more common
in high redshift radio galaxies, therefore they may help us in better
understanding these far-away systems.

\begin{acknowledgements}
We would like to thank the referee, Jacqueline van Gorkom, for her
valuable (and prompt) comments.  The WSRT is operated by the ASTRON
(Netherlands Foundation for Research in Astronomy) with support from
the Netherlands Foundation for Scientific Research (NWO).  The VLA is
a facility of the National Radio Astronomy Observatory, which is
operated by Associated Universities Inc., under cooperative agreement
with the National Science Foundation.
\end{acknowledgements}


\begin{thebibliography}{}
  
\bibitem[]{} Barnes J.E. 2002, MNRAS submitted  (astro-ph/0110581)
\bibitem[]{} Briggs D. 1995, PhD thesis, New Mexico Institute of Mining and
  Technology
\bibitem[]{} Capetti A., de Ruiter H.R., Fanti R., Morganti R., Parma P. \& Ulrich
M.H.  2000, A\&A 362, 871
\bibitem[]{} Ebneter K., 1989, PhD Thesis, Univ.  California, Berkeley;
\bibitem[]{} Evans A.S., Kim D.C., Mazzarella J.M., Scoville N.Z. \& Sanders D.B.  1999a, ApJ 521, L107
\bibitem[]{} Evans A.S., Sanders D.B., Surace J.A. \& Mazzarella J.M.  1999b, ApJ 511, 730
\bibitem[]{}   Falco, E., Kurtz, M., Gellar, M. et al. 2000, The Updated Zwicky
  Catalog (UZC), vol. 1 p. 1 based on 1999 PASP 111,438
\bibitem[]{} Fanti C., Fanti R., de Ruiter H.R. \& Parma P. 1987 A\&AS 69, 57
%\bibitem[]{} Giovannini G., Cotton W.D., Feretti L., Lara L. \& Venturi T. 2001, ApJ 552, 508
\bibitem[]{} Heckman T.M., Smith E.P., Baum S.A. et al. 1986 ApJ, 311, 526
\bibitem[]{} Heisler C.A., Norris R.P., Jauncey D.L., Reynolds J.E. \& King
  E.A. 1998, MNRAS 300, 1111
\bibitem[]{} Hibbard J.E. 1995, PhD Thesis Columbia University
%\bibitem[]{} Hibbard J.E. \& Mihos J.C. 1995b, AJ, 110, 140
\bibitem[]{} Hibbard J.E. \& van Gorkom J.H. 1996, AJ 111, 655
\bibitem[]{} Hibbard J.E., van der Hulst J.M., Barnes J.E. \& Rich R.M. 2001, 122, 2969
\bibitem[]{} Hibbard J.E., van Gorkom J.H., Rupen M.P. \& Schiminovich D., 2002 in
in "Gas \& Galaxy Evolution", J.E.  Hibbard, M.P.  Rupen \& J.H.  van Gorkom (eds.), also available at  http://www.nrao.edu/astrores/HIrogues/
%\bibitem[]{} Kennicutt R.C. 1989, ApJ 344, 685 
\bibitem[]{} Kormendy J. \& Richstone D. 1995, ARA\&A 33, 581
\bibitem[]{} Lees J.F. 1994, in {\sl Mass-Transfer Induced Activity in
Galaxies''}, ed.  Shlosman I., Cambridge Uni.  Press, p.  432 (NGC4278)
\bibitem[]{} Lim J. \&  Ho P.T. 1999 ApJ 510,L7
\bibitem[]{} Lim J., Leon S. \& Combes F. 2000, ApJ 545, 93 
\bibitem[]{} Mazzarella J.M., Graham J.R., Sanders D.B. \&
 Djorgovski S. 1993, ApJ 409, 170;
\bibitem[]{} Mihos J.C. \& Hernquist L. 1994, ApJ 431, L9
\bibitem[]{} Morganti R., Sadler E., Oosterloo T., Pizzella A. \& Bertola F.,
  1997, AJ, 113, 937 
\bibitem[]{} Morganti R., Oosterloo T., Tadhunter C.N., van Moorsel, Killeen
N. \& Wills
K.A.  2001, MNRAS 323, 331  
\bibitem[]{} Morganti R., Oosterloo T.A., Tinti S., Tadhunter C.N., Wills
K.A. \& van Moorsel G.  2002a, A\&A 387, 830
\bibitem[]{} Morganti R., Oosterloo T.A., Tinti S., Tadhunter C.N., Wills
K.A. \& van Moorsel G.  2002b,  in {\sl Seeing through the dust}, R.  Taylor, T.
Landecker, A.  Willis, ASP in press (astro-ph/0112269)
\bibitem[]{} Morganti R., Tadhunter C., Oosterloo T., Holt J., Tzioumis A. \&
Wills K. 2003, PASA in press (astro-ph/0212321)
\bibitem[]{} O'Dea C.P. 1998, PASP 110,493
\bibitem[]{} Oosterloo T.A., Morganti R., Sadler E.M.  1999, in {\sl Star
Formation in Early-Type Galaxies}, eds. P.Carral \& J.Cepa, ASP Conf.Proc. p.84
\bibitem[]{} Oosterloo T., Morganti R., Vergani D., Sadler E.M. \& Caldwell N.
  2002a, AJ, 123, 729
\bibitem[]{} Oosterloo T.A., Morganti R. \& Sadler E.M., 2002b, in "Gas \& Galaxy
  Evolution", J.E.  Hibbard,
M.P.  Rupen \& J.H.  van Gorkom (eds.) p.251 (astro-ph/0009112)
\bibitem[]{} Pihlstr\"om Y.M. 2001, PhD. Thesis, Chalmers University \& Onsala
Observatory
\bibitem[]{} Raimond E., Faber S.M., Gallagher J.S. \& Knapp G.R. 1981, ApJ, 246, 708
\bibitem[]{} de Ruiter, Parma P., Fanti R., Capetti A. \& Morganti R.  2002,
A\&A in press (astro-ph/021050)
\bibitem[]{} Sanders D.B. \& Mirabel I.F. 1996, ARA\&A 34, 749
\bibitem[]{} Sault, R.J., Teuben, P.J. \& Wright, M.C.H. 1995, in Astronomical
  Data Analysis Software and Systems IV, R. Shaw, H.E. Payne and J.J.E. Hayes,
  eds, Astronomical Society of the Pacific Conference Series, 77, p. 433
\bibitem[]{} Schiminovich D., van Gorkom J.H., van der Hulst J.M. \& Kasow S., 1994, ApJ, 423, L101
\bibitem[]{} Schiminovich D., van Gorkom J.H., van der Hulst J.M. \& Malin D. 1995, ApJ 444, L77
\bibitem[]{} Smith E.P.\& Heckman T.M.  1989, ApJ, 341, 658; 
\bibitem[]{} Tadhunter C.N., Wills K.A., Morganti R., Oosterloo 
T. \& Dickson R. 2001, MNRAS 327, 227 
\bibitem[]{} Tadhunter C.N., Dickson R., Morganti R., Robinson T.G., Wills K. \& Villar-Martin M., Hughes M. 2002, MNRAS 330, 977
\bibitem[]{} Tingay S.J., Jauncey D.L., Reynolds J.E., Tzioumis A.K., King
E.A. et al. 1997, AJ 113, 2025
\bibitem[]{} Vader J.P., Frogel J.A., Terndrup D.M. \& Heisler C.A. 1993 AJ 106, 1743
\bibitem[]{} van Gorkom J.H., Knapp G.R., Raimond E., Faber S.M. \& Gallagher J.S.  1986, AJ, 91, 791
\bibitem[]{} van Gorkom J.H., van der Hulst J.M., Haschick A.D. \& Tubbs A.D. 1990, AJ 99, 1781
\bibitem[]{} van Gorkom J.H. \& Schiminovich D. 1997, in {\sl The Nature of Elliptical Galaxies},  ed. M.  Arnaboldi; G. S. Da Costa; and P. Saha (1997), ASP Conference
  Series; Vol. 116, p.310
\bibitem[]{} van Ojik R., Roettgering H.J.A., Miley G.K. \&
 Hunstead R.W. 1997, A\&A 317, 358
%\bibitem[]{} Verdoes Kleijn G.A., Baum S.A., de Zeeuw
%  P.T. \& O'Dea C.P. 1999, AJ 118, 2592
\bibitem[]{} Verdoes Kleijn G.A., Baum S., de Zeeuw T.P. \& O'Dea C.P. 2002a, AJ 123, 1334
\bibitem[]{} Verdoes Kleijn G.A. 2002b, PhD Thesis, Leiden University 
\bibitem[]{} V\'eron-Cetty M.-P., Woltjer L., Ekers R.D. \& Staveley-Smith L. 1995, A\&A 297, L79
\bibitem[]{} Villar-Marti\'n M., Vernet J., di Serego Alighieri S., Fosbury
R., Pentericci L., Cohen M., Goodrich R. \& Humphrey A.  2002, MNRAS in press
(astro-ph/0206118)
\bibitem[]{} Wills K.A., Tadhunter C.N., Robinson T.G. \& Morganti R.  2002, MNRAS 333, 211
\bibitem[]{} Wilson A.  1996, in {\sl `` Energy Transport in radio galaxies and
quasars''}, Hardee, Bridle, Zensus eds., ASP Conf.  Series p.9

\end{thebibliography}
\end{document}